\documentclass{osa-article}

\journal{oe}

\articletype{Research Article}
\usepackage{siunitx}
\graphicspath{{figs/}}  
\newcommand{\ion}{$^{3+}$}
\newcommand{\um}{\si{\micro\m}~}
\DeclareUnicodeCharacter{2009}{\,}	
\usepackage{lineno}

\begin{document}

\title{Diode pumped silicate fiber for visible laser emission}

\author{Matthew R. Majewski,\authormark{1,*} and Stuart D. Jackson\authormark{1}}

\address{\authormark{1}MQ Photonics Research Centre, Faculty of Science and Engineering, Macquarie University, Sydney, NSW 2109, Australia}

\email{\authormark{*}matthew.majewski@mq.edu.au} 



\begin{abstract}
We demonstrate yellow laser emission using silicate glass fiber as the gain medium. 
By employing core pumping using widely available GaN diode lasers with emission at 445~nm, we show that Dy\ion-doped aluminosilicate glass fiber can be readily excited creating sufficient gain at 581~nm. 
In this proof of concept demonstration, the maximum output power generated was 3~mW with a slope efficiency approximately 1.8\% with respect to injected pump. 
This first in class demonstration opens a new field within fiber laser research.
\end{abstract}

Fiber lasers have been shown to emit light efficiently across 3 octaves from approximately 0.3~\um\cite{El-Agmy2008} to 4~\um\cite{Schneider1995a}. 
Every high (>100~W) power demonstration across this range has involved silicate glass fiber. 
At the spectral limits of fiber laser emission, low phonon energy glasses, typically fluoride glass, instead of the more common silicates are used. 
The relatively low phonon energy associated with fluoride glass reduces quenching from multiphonon decay and results in sufficiently long upper laser level lifetimes of visible and mid-infrared transitions. 
Combined with the demonstration of high slope efficiencies using a judicious choice of the pump processes, rare earth doped fluoride glass fibre has created enormous opportunities in applications where either a short or relatively long wavelength is needed.

There are important limitations to the use of fluoride fiber. 
Fluoride glass fiber is more expensive and comparatively difficult to post process compared to fibers made from compositions located within the silicate glass family. 
Fluoride glass has a transition temperature of only 260~C, which also limits the thermal load under high power pumping. 
As a result, lasers using silicate glass fiber emit output power levels at least two orders of magnitude higher than those lasers using fluoride glass fiber~\cite{Xiao2012,Jeong2009,Aydn2018}.
The range of silicate fiber-based componentry e.g., couplers, wavelength-division-multiplexers, and beam combiners is comparatively much larger, making any number of stable, reliable, and versatile fiber laser arrangements possible. In contrast, there are only a handful of fluoride glass components demonstrated and even fewer available commercially.

While the choice of fluoride glass as the host medium for laser emission in the mid-infrared is well established because mid-infrared laser transitions occur between closely-spaced energy levels meaning phonon-induced quenching is always a concern, this is not universally true for visible laser emission. 
Visible light comprises of photons with energies many times higher than the maximum phonon energies of even the most phonon energetic glasses, e.g., the phosphates.
However, as the photon energy increases, for some transitions multiphonon quenching becomes a concern because of the presence of intermediate energy levels between the upper and lower laser levels; for most visible laser transitions, low phonon energy glasses have been used.
Examples include blue emission from Tm\ion~ where low phonon energy glass enables an upconversion pump process~\cite{Laperle2000,Duclos1995}, and red emission from Pr\ion~ where direct multiphonon quenching of the upper laser level is reduced sufficiently to allow oscillation~\cite{Kajikawa2016,Fujimoto2019,Lord2021}.
To date, only Sm\ion~ has been shown to have an energy level structure which allows for visible laser oscillation in silicate glass~\cite{Farries1988}. 
In this system, a centre wavelength of 651~nm was produced after pumping at 488~nm using an argon ion laser.

The energy level structure of Dy\ion~ is also favorable for visible laser emission using a silicate glass host. 
While the yellow emitting $^4F_{9/2}\rightarrow^6H_{13/2}$ transition has a number of intermediate energy levels, the upper laser level is separated from its adjacent lower energy level by approximately \SI{7500}{\per\cm}.
This is many times the nominal maximum phonon energy of \SI{1100}{\per\cm} associated with silicate glass.
We therefore show for the first time, in a simple proof-of-concept demonstration, that visible fiber laser emission from the yellow transition of Dy\ion~ is possible using high phonon energy glasses such as the silicates. 
Though our measured performance is not at the level achieved with state-of-the-art fluoride fibre, it nevertheless shows the potential of silicate glass as the ideal gain medium for high power laser emission in the yellow region of the spectrum. 

\begin{figure}[!htbp]
	\centering
	\includegraphics{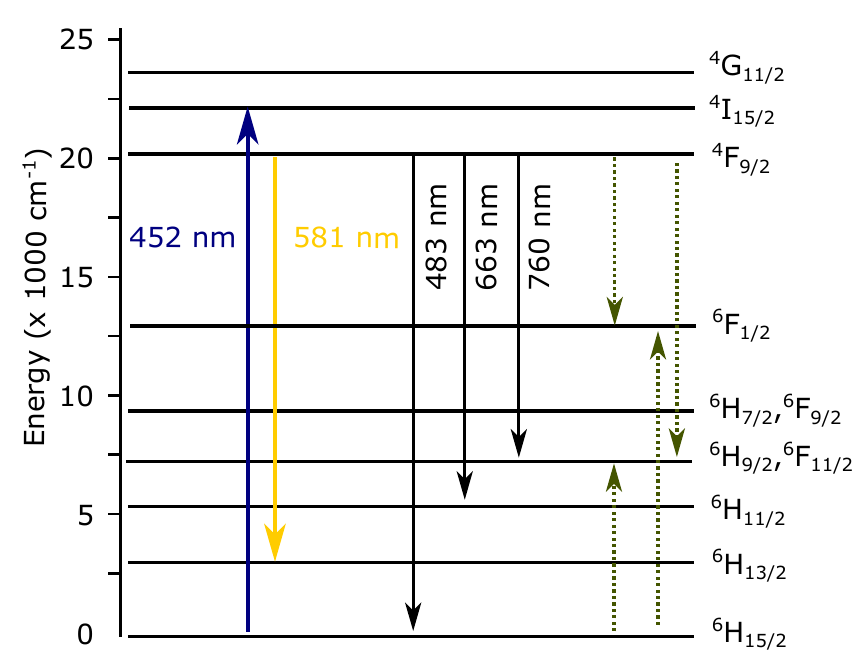}
	\caption{Simplified energy level diagram of Dy\ion: solid arrows indicate pump and radiative transitions, while dashed arrows show possible cross-relaxation routes.}
	\label{fig:energy_level}
\end{figure}

An energy level diagram illustrating the relevant transitions is shown in Fig.~\ref{fig:energy_level}.
Pumping is provided by ground state absorption in the blue to $^4I_{15/2}$ followed by fast non-radiative decay to $^4F_{9/2}$.
From this metastable level there are several possible radiative transitions, dominated by the chosen laser oscillation line around 580~nm. 
Laser action from this transition was demonstrated more than two decades ago in fluoride glass (ZBLAN) fiber~\cite{Limpert2000}, yet suffered both from low quality fiber and the reliance on an Ar ion laser as the pump source.
Recently, ZBLAN fiber lasers operating on this transition are receiving a resurgence of interest because of increased fiber quality and the widespread availability of GaN laser diodes emitting around 450~nm.
This has led to recent demonstrations of both high efficiency~\cite{Wang2019b} and Watt-level output power~\cite{Zou2021}.

For this simple proof-of-concept demonstration, we employ an aluminosilicate glass fibre with a Dy\ion~ concentration of 3000~ppm molar, a core diameter of 25~\um and an NA of 0.18 (IxBlue, France). 
The dopant concentration was chosen to be relatively low. 
It has been shown~\cite{Nagli1997,Sun2010,Liu2020,Bowman2012}  that multiple cross-relaxation processes (CR), as depicted in Fig.~\ref{fig:energy_level}, can quench the $^4F_{9/2}$ upper laser level significantly. 
While a Dy\ion~ concentration of 2000~ppm molar has been shown to inhibit significant CR in fluoride glass, it is not clear what the upper bound for the concentration without significant CR will be in silicate glass. 
To be cautious, we chose 3000 ppm molar to provide relatively strong core pumping absorption due to concern about background loss resulting from primarily from Rayleigh scattering. 
As is well known, scattering is typically higher in the visible region of the spectrum in silicate glass compared to fluoride glass; a feature made worse when more dopants are added. 
The large NA was a consequence of adding Al$_2$O$_3$ modifier to the glass. 
We calculate a single mode cutoff for this fiber of 5.8~\um, resulting in highly multimode operation at both pump and laser wavelengths.
The large core diameter allowed us to core pump the fiber with a reasonable launch efficiency using a simple imaging setup of the output from a single high power, low brightness, GaN diode laser (Osram).

Before we assessed the laser performance, we carried out a basic spectroscopic analysis of the fiber.
First, we measured, using a previously described frequency domain method~\cite{Majewski2018b} the $^4F_{9/2}$ energy level lifetime as a function of pump power. 
When extrapolated to zero pump power and for sufficient modulation frequency, this method gives a reliable measurement of the lifetime which we determined to be approximately \SI{500}{\micro\second}. 
Compared to the \SI{1500}{\micro\second} lifetime reported in the literature for a Judd-Ofelt calculation of silicate~\cite{Nagli1997}, it is likely that a quenching process is present. 
We estimate from an energy gap law calculation that the contribution to the total decay of this level from multiphonon emission is only \SI{1.4e3} (i.e., the multiphonon decay rate is \SI{7e-4}{\per\second}), further suggesting cross-relaxation is the dominant lifetime reduction mechanism.
Given that aluminosilicate glasses have stronger inhomogeneous broadening relative to ZBLAN, it is possible that the rates of CR may be higher in this silicate glass composition compared to ZBLAN that results in significant lifetime quenching even at this moderate doping level. 
The lifetime dependence of the $^4F_{9/2}$ energy level as a function of the Dy\ion~ concentration will be the examined in detail in a future spectroscopic investigation. 

\begin{figure}[!htbp]
	\centering
	\includegraphics{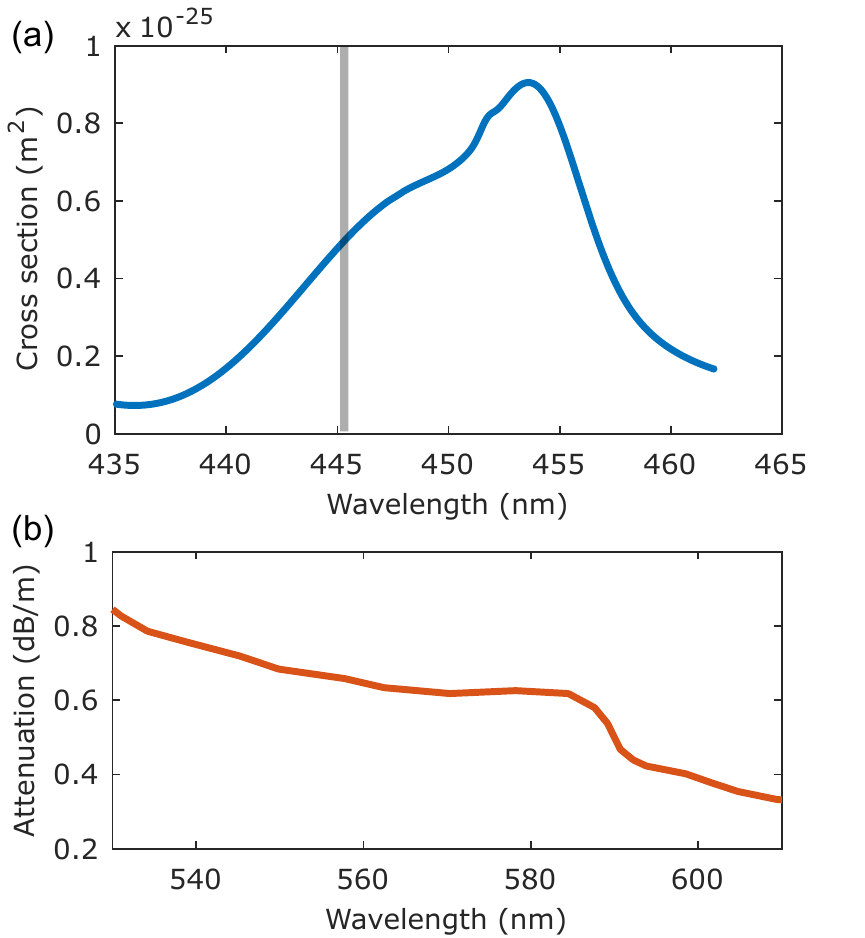}
	\caption{a) Pump absorption cross section estimation from manufacturer provided cutback data, with our pump wavelength indicated and, b) attenuation data around the laser wavelength.}
	\label{fig:Ixdata}
\end{figure}

Figure \ref{fig:Ixdata}~shows the absorption spectrum of the fiber at the pump and lasing regions which we have extracted from data supplied from the fibre manufacturer. 
The absorption cross sections in Fig.~\ref{fig:Ixdata}(a) are estimated by subtracting the shape of absorption peaks reported in bulk germanosilicate glass~\cite{Liu2020} from raw cutback attenuation date.  
At the pump wavelength of 445 nm, the absorption cross section is nominally \SI{5e-26}{\meter\squared}, which translates to a core absorption coefficient of 28~dB/m. 
We were not able to temperature tune the pump laser without significant degradation to the available pump power, so all our measurements involved an extrinsic absorption by Dy\ion~  of approximately half of the peak value. 
Note that it is difficult to obtain GaN diodes that precisely match with the peak in the absorption. 
Most suppliers supply commercial-grade GaN diodes across a fairly broad wavelength range (typically >10~nm). 
We selected the diode used after measuring the emission from six diodes. 
Despite this being clearly sub-optimal, it was sufficient for this proof-of-concept demonstration. 
The intrinsic background absorption (likely arising from Rayleigh scattering) was 7~dB/m which is more than three times higher compared to ZBLAN glass. 
It is well known that the addition of modifiers and active ions to pure silica glass, which has a nominal scattering coefficient of <50 dB/km in this wavelength region, increases glass inhomogeneity significantly. 
Note that the fibre was manufactured using the solution doping technique developed for high power infrared emission. 

\begin{figure}[!htbp]
	\centering
	\includegraphics{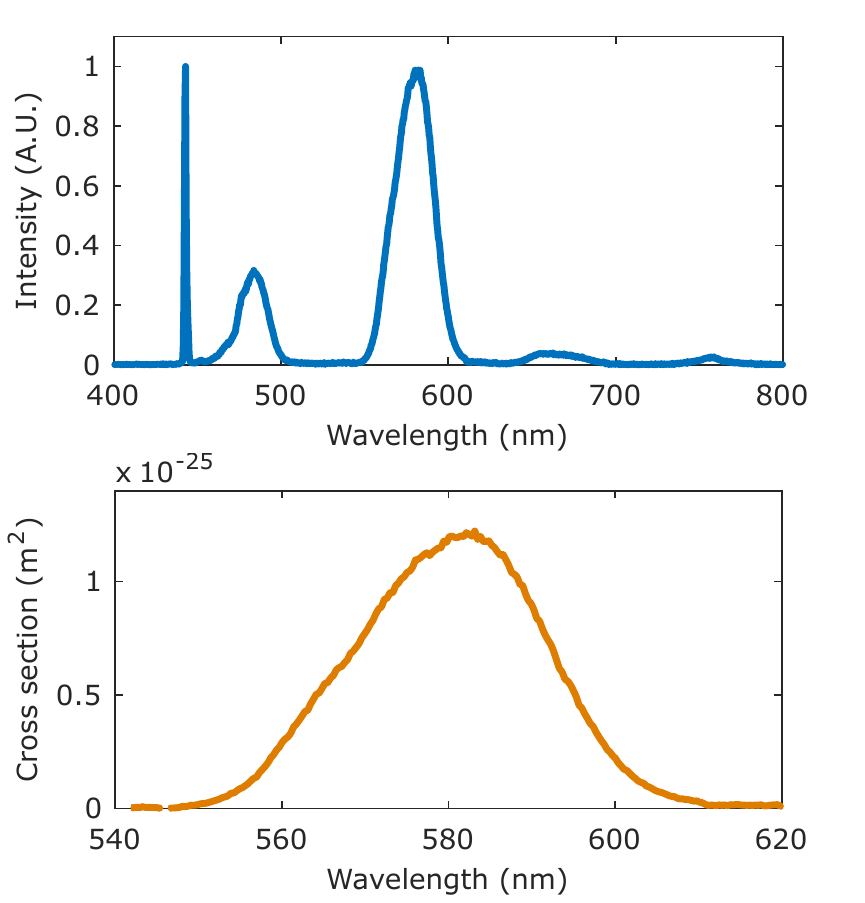}
	\caption{a)Measured side light fluorescence spectrum including scattered pump light and, b) calculated emission cross section spectrum.}
	\label{fig:fl}
\end{figure}

The measured side light fluorescence spectrum from the fiber under excitation at 445~nm is shown in Fig.~\ref{fig:fl}(a). 
We used this measurement to estimate the branching ratios by spectral integration, which yields a value for $^4F_{9/2}\rightarrow^6H_{13/2}$ of 0.77.
This is in fair agreement with previously reported values for bulk silicate of 0.72 and 0.71, calculated and measured respectively~\cite{Nagli1997}.
The large branching ratio is encouraging and should provide for low to moderate pump power levels at threshold. 
Fig.~\ref{fig:fl}(b) presents the result of the application of the F\"uchtbauer--Ladenburg equation to the fluorescence spectrum, using literature-sourced spectroscopic values. 
The peak emission cross section of \SI{1.2e25}{\meter\squared} occurs at 581~nm, which is red shifted by approximately 8~nm relative to ZBLAN glass~\cite{Zou2021}. 
It is important to note that the cross section at 589~nm, an important wavelength for guide star applications, is approximately 82\% of the peak value.

\begin{figure}[!htbp]
	\centering
	\includegraphics{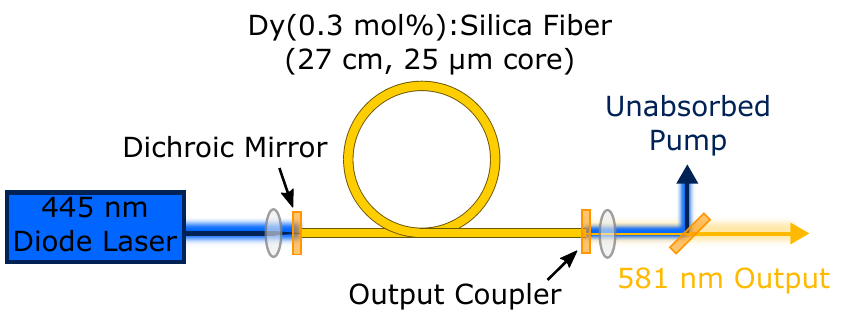}
	\caption{Fiber laser cavity schematic}
	\label{fig:cavity}
\end{figure}

To determine the viability of silicate glass as a suitable host for yellow laser emission using the Dy\ion~, the performance under moderate CW pumping must be assessed. 
A basic experimental setup is shown in Fig.~\ref{fig:cavity}. 
Commercial grade GaN diodes capable of up to 2.6~W at 2.1~A drive current were used. 
The output from the diode laser was collimated and then focussed into the core of the fiber using 12~mm and 20~mm focal length lenses, respectively. 
We estimate the launch efficiency to be approximately 27\%; the highest we could acheive with this simple two-lens circularly symmetric arrangement. 
The spot size at the focal plane was estimated to be 57~\um x \SI{6}{\micro\meter}. 
This beam expansion configuration represented the best tradeoff between loss due to spatial overfill in the slow axis and angles exceeding the NA in the fast axis.
Improved launch efficiency should readily be realized in future experiments designed to optimise the performance with a more sophisticated asymmetrical imaging system.
Two cavity mirror sets were used. 
Each employed a pump input mirror that was >99\% reflecting at 581~nm and 98\% transmitting at the pump. 
%
%
%
Laser output mirrors with 5\% and 11\% transmission around 581~nm formed the output couplers for the experiments. 
The output was collimated and measured with a calibrated Si photodiode power detector, and the unabsorbed pump was rejected with a longpass filter; laser output powers are corrected for slight losses from both the lens and filter.  

\begin{figure}[!htbp]
	\centering
	\includegraphics{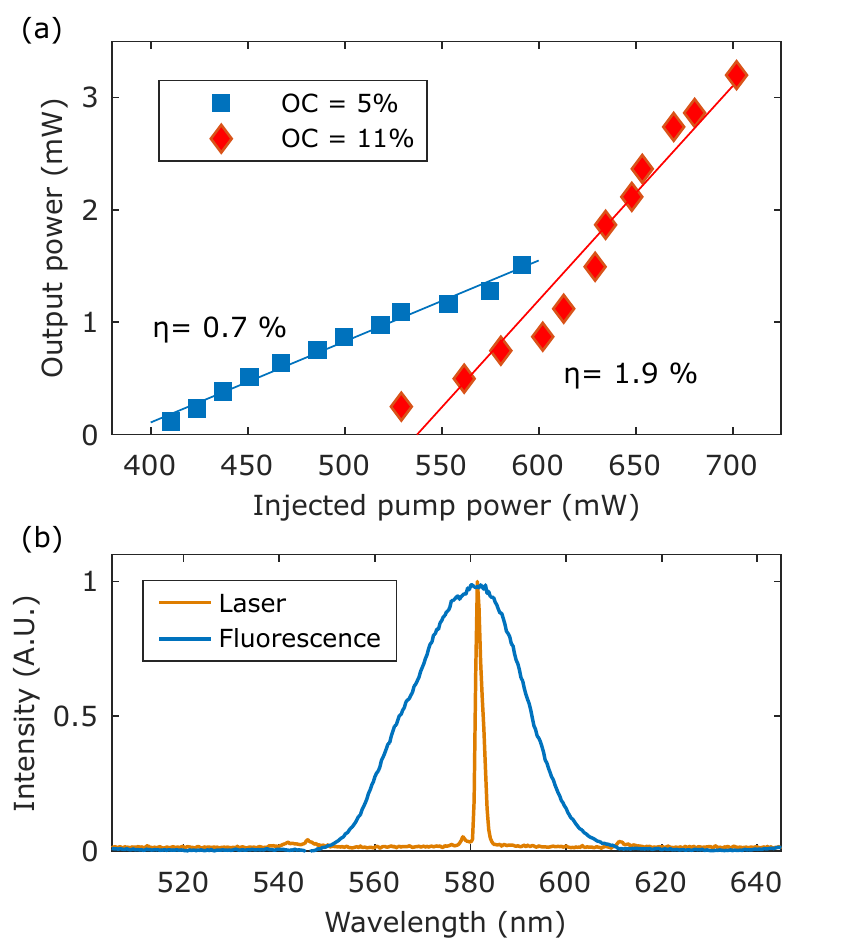}
	\caption{a) Laser output as a function of pump power with slope efficiencies ($\eta$) indicated and, b) laser optical spectrum overlayed with the fluorescence spectrum.}
	\label{fig:dy_performance}
\end{figure}

Figure~\ref{fig:dy_performance}(a) shows the fundamental characteristics of the output. 
The lowest threshold pump power was 400~mW injected, and the maximum output power of 3~mW was achieved at a slope efficiency of 1.9\%. 
There was no evidence that ASE at other wavelengths contributed significantly to the output. 
Note there were no corrections made to consider intrinsic loss of the pump power which if corrected, would boost the slope efficiency well beyond 2\%. 
It is clear that at this very early state of development, the silicate glass fibre seriously under performs relative to fluoride fibre. 
We must stress that for this proof-of-concept experiment that no special effort was made to create silicate fibre that had a particularly high transmission in the visible region of the spectrum. 

Both the output spectrum optical spectrum and the fluorescence spectrum are seen in Fig.~\ref{fig:dy_performance}(a). 
The output emission wavelength overlaps with the peak fluorescence level, in contrast to the red-shifted output observed with ZBLAN systems~\cite{Limpert2000}, evidence that this laser is operating as a true four level system. 
In ZBLAN glass, the lower laser level has a lifetime of \SI{600}{\micro\second} which is approaching half the lifetime of the upper laser level.
As a result, there can exist a sizeable population in the lower level which forces the emission wavelength to red shift relative to the peak fluorescence. 
In aluminosilicate glass, the lower laser level is nonradiatively quenched entirely.

The low slope efficiency achieved can be attributed in large part to a few key factors.
First results from examination of the absorption spectrum of the fibre seen in Fig.~\ref{fig:Ixdata}. 
We estimate that the background loss at 445~nm and 581~nm is 7 dB/m and 0.7 dB/m, respectively. 
For the fiber length of 27~cm that was used, we estimate that the intrinsic pump loss resulting from scattering comprises approximately 40\% of the total pump absorption. 
Given the large intrinsic fiber loss at this wavelength, the blue-shifted pump wavelength relatively to the peak contributed to the overall lower than expected performance of the fiber laser. 
One key issue associated with the acquisition of the blue-emitting GaN diode laser is wavelength selectivity.
Blue diode lasers are primarily fabricated for the domestic consumer market whereby precision in the emission wavelength is not a key concern. 
While the need in the domestic consumer market has driven the cost/W parameter to exceptionally low values, the need for wavelength precision will grow as the demand for GaN diode laser for optical pumping applications grows. 
It is demonstrations such the current one that drive the need for more precision in the available choices of wavelength from commercial GaN diode lasers.

The intrinsic glass loss at 581~nm, while being lower than the pump loss, also contributed significantly to the overall poor performance of the fiber laser relatively to systems employing fluoride fibre.  
The problem was made worse by the large number of round trips within the resonator for the mirror sets that were tested. 
We expect much higher slope efficiencies for higher ratios of output coupling relative to background loss, though this will increase oscillation threshold. 
For the present proof-of-concept experiment, we limited the experiment to a single diode laser and consequently, the available pump power and the range of output coupling values was restricted. 
Future experiments will involve multiplexing the pump light to increase the available pump power using the fact that the diode laser emission from GaN is polarised and, that the fibre can also be pumped from each end. 

It is clear that diode pumping Dy\ion~-doped silicate glass fibre has a long way to go in order to match the current performance records established using fluoride fibre. 
But our proof-of-concept demonstration shows that the same glass composition that is used for all the demonstrations of kW-class fiber laser emission in the infrared is also useful for fibre lasers systems developed for yellow laser emission. 
While it will be necessary to make significant changes to the fibre fabrication process to create rare earth doped aluminosilicate glass with much better transmission in the visible region, we have nevertheless shown, along with the only other demonstration over three decades ago, the potential of rare earth doped silicate glass fibre for visible light generation.

In summary, we have demonstrated a new class of diode pumped fiber laser. 
While the output performance was poor relative to reported demonstrations using fluoride glass fibre, with a maximum output power of only 3 mW achieved, we have re-awakened the application of rare earth doped silicate glass for visible light generation. 
Importantly, with the strong quenching of the lower laser level and the output power potential of silicate glass relative to fluoride glass, future 100~W level emission at Stokes-limited efficiencies is a strong possibility.  
Given that fiber gratings are easily fabricated into silicate fiber and that the fluorescence spectrum extends well beyond 589 nm, applications requiring guide-star sources may in the future adopt Dy\ion~-doped silicate glass fibre as the primary gain medium.

	\section*{Funding}
	Asian Office of Aerospace RD (FA2386-19-1-0043).

	\bigskip
	\noindent\textbf{Disclosures.} The authors declare no conflicts of interest.

	\bibliography{library}

\end{document}